\documentclass{article}
\usepackage[utf8]{inputenc}
\usepackage[english]{babel}
\usepackage[a4paper,top=3cm,bottom=3cm,left=3cm,right=3cm,marginparwidth=3cm]{geometry}
\usepackage[english]{babel}
\usepackage[nottoc]{tocbibind}
\usepackage{rotating}
\usepackage{graphicx}
\usepackage[hyphens]{url}
\usepackage{hyperref}
\hypersetup{breaklinks=true}
\graphicspath{ {./images/} }
\usepackage{tabularx}
\usepackage{blindtext}
\usepackage{multicol}
\usepackage[numbers]{natbib}
\AtBeginDocument{\AtBeginShipoutNext{\AtBeginShipoutDiscard}}

\begin{document}
\title{Refugee status determination: how cooperation with machine learning tools can lead to more justice}
\author{Claire Barale \footnote{Claire Barale, claire.barale@ed.ac.uk, School of Informatics, The University of Edinburgh, Informatics Forum, 10 Crichton street, Edinburgh EH8 9AB, UK}}
\date{}

\begin{flushleft}
\maketitle
\vspace{-1\baselineskip}
            \hrule
\end{flushleft}

\begin{multicols}{2}

\setcounter{page}{1}
\section*{Abstract}
According to the UNHCR data, approximately 26.6 millions of people were refugees and 4.4 million were asylum seekers as of mid-2021. At the core of this global crisis is the procedure of refugee status determination, that is to say deciding who is granted refugee status and who is not. \par

Previous research on refugee status adjudications has shown that prediction of the outcome of an application can be derived from very few features with satisfactory accuracy. Recent research work has achieved between 70 and 90\% accuracy using text analytics on various legal fields among which refugee status determination. Some studies report predictions derived from the judge identity only. Additionally most features used for prediction are non-substantive and external features ranging from news reports, date and time of the hearing or weather. On the other hand, literature shows that noise is ubiquitous in human judgments and significantly affects the outcome of decisions. It has been demonstrated that noise is a significant factor impacting legal decisions. We use the term ``noise" in the sense described by D. Kahneman, as a measure of how human beings are unavoidably influenced by external factors when making a decision. In the context of refugee status determination, it means for instance that two judges would take different decisions when presented with the same application. \par

This article explores ways that machine learning can help reduce noise in refugee law decision making. We are not suggesting that this proposed methodology should be exclusive from other approaches to improve decisions such as training of decision makers, skills acquisition or judgment aggregation, but rather that it is a path worth exploring. We investigate how artificial intelligence and specifically data-driven applications can be used to benefit all parties involved in refugee status adjudications. We specifically look at decisions taken in Canada and in the United States. Our research aims at reducing arbitrariness and unfairness that derive from noisy decisions, based on the assumption that if two cases or applications are alike they should be treated in the same way and induce the same outcome. \par
After identifying areas of refugee status adjudications subject to noise, we explain potential benefits of legal prediction and models that help classify cases and evaluate alternatives based on a similarity analysis. Based on the current state of the art, we can exploit the capacity of machine learning to synthesize databases' content for prediction, classification and information retrieval purposes. We refer for instance to the experiment lead on Refworld, a UNHCR maintained database that allows users to quickly find related cases and relevant legislation. While we agree that algorithms are not destined to replace human decision making in refugee status determination, we show that cooperation between human judgment and AI attributes can orient the intuition of the decision maker toward fairer decisions, that benefit both legal professionals and asylum seekers.

\vspace{\baselineskip}
            \hrule

\section{Background}

\subsection{Refugee status determination}
Refugee law is a part of international law and is closely linked to human rights law. It mainly relies on two legally binding texts of universal application: the 1951 Refugee Convention \cite{unhcr_convention_1951} and the 1967 Protocol \cite{un_protocol_1967}. To be granted refugee status, an applicant needs to demonstrate a “well-founded fear of being persecuted" for one of the following reasons: ``race, religion, nationality, membership of a particular social group or political opinion" (art. 1A(2) of the Convention, \cite{unhcr_convention_1951}). The decision is taken by a judge based on interviews (usually conducted by civil servants), hearings, and written applications. Applications are often prepared with the help of case workers or refugee claim officers, NGOs, sometimes but not necessarily lawyers. The most important evaluation criterion of an application is arguably the credibility assessment. Claims are then compared to country reports that are regularly updated by the United Nations High Commissioner for Refugees (UNHCR) and NGOs such as Human Rights Watch or Amnesty International.

\subsection{Legal AI and refugee law}
AI-powered legal systems are not a new idea and have been researched as early as the 50s \cite{allenlayman_symbolic_1956} and used since the 70s in the form of expert systems \cite{buchananheadrick_speculation_1970, popple_pragmatic_1996}, based on symbolic knowledge representation and rules. Data-driven applications have been used in border control and migrations whether by states, international organizations as the UNHCR or the European Union and by NGOs, for a range of tasks such as prediction of migration flows, background checks, visa applications, legal information mining, or face recognition \cite{molnar_botsgate_2018, european_commission_directorate_general_for_migration_and_home_affairs_opportunities_2020}. \par

Starting 2019, the UNHCR conducted an information retrieval project focusing on easier navigation and document search on \textit{RefWorld} database \cite{service_unhcrmediumgiving_2020}. \textit{RefWorld} is a refugee law database maintained by the UN, which indexes cases, conventions, relevant legislation, policy documents, country reports, and news. It is widely used by lawyers and counsels to write applications and by judges in their decision-making process. New proposed functionalities include extracting citations, finding related cases and decisions, and smart-searching for policy documents and legislation.   \par

Introducing data-driven models in the legal domain commonly triggers a series of difficulties including: technical constraints (narrow AI and need for manual engineering), sparse data and uncertainty, balance between human and machine legal reasoning, risk of unjustified and unfair decisions, lack of information and support for the asylum seeker, reinforcement of existing biases, interpretability and accountability, privacy concerns and impact of the use of AI on the legal process and the law. \par

Specific literature studying international law and AI, and particularly refugee law and AI, is sparse although one can rely on a larger field of research if considering legal AI in general.

\subsection{Predictive analysis}
We are specifically interested in legal prediction, that often amounts to :
\begin{enumerate}
    \item quantify legal uncertainty i.e. for instance predict the risk of an unjustified decision, as studied by \citeauthor{cameron_artificial_2021} for asylum decisions
    \item predict the outcome of a case \cite{dunn_early_2017, chen_asylum_2017}
\end{enumerate}

Previous  results  show  that  different  machine  learning  methods  have  been  successfully used  for  legal  prediction with  satisfactory  accuracy  levels.
Recent research has achieved a high level of accuracy using text analytics and improved algorithms. Experiments have been made comparing machine learning models (support vector machines, logR, convolutional neural networks, recurrent neural network) on legal data sets gathering decisions from the Supreme Court of the United-States \cite{katz_general_2017,martin_competingsupreme_2004,ruger_supremeforecast_2004,undavia_comparative_2018} or the European Court of human rights \cite{aletras_predicting_2016, medvedeva2020using, kaur2019convolutional}, results of which can be easily extended to asylum decisions. Similar studies have been conducted on asylum decisions data sets \cite{chen_asylum_2017, dunn_early_2017, rehaag2012judicial}.

\subsection{Noise in legal decision making}
In this paper, in addition to AI and legal decision making, we also draw insights from cognitive sciences and specifically cognitive psychology. Interestingly, asylum decisions have been studied as a paradigmatic example of noisy decision by \citeauthor{noisekahneman}.

In statistics, noise refers to a random irregularity within a sample. Noise in decision making is defined as an ``unwanted variability" \cite{sunstein2021governing} that has consequences as it can produce errors of judgment and inconsistency in the outcome of the decisions. Noise is is difficult to detect and correct as there are no recognizable patterns in errors (unlike errors that derive from bias). Three kinds of noise can be distinguished: 
\begin{enumerate}
    \item  ``Occasion noise" -- decisions can be subject to external factors such as the time of the day or the place of the hearing for instance
    \item ``Level noise" -- different decision makers can render different decisions on the exact same case
    \item ``Pattern noise" -- different decision makers are diversely influenced in their decision by different factors
\end{enumerate}

Definitions of variability in decision making distinguish two broader categories of noise: interpersonal and intrapersonal noise. The latter refers to the potential impact that external features have on an adjudicator's perception and judgment of a case (occasion noise). Interpersonal noise on the other hand regroups both pattern and level noise and refers to the difference of the judgment that is found between two different adjudicators.

While biases have been widely researched both in legal decisions and in AI, it is important that it is distinguished from noise. Biases are usually defined as systematic errors for which it is possible to identify a pattern, whether the bias is algorithmic or due to human cognition. Decision making is often both noisy and biased at the same time. \par

This paper focuses on studying and proposing methods to prevent noise in asylum decisions. Indeed, some argue that algorithmic decision making is a solution to the problem of noise in judgments as algorithms have the capacity to eliminate noise completely \cite{sunstein2021governing}. However, it is common knowledge that algorithms are unable to render bias-free decisions. In many cases, computer application can in fact reinforce biases found whether in data, algorithm architecture or outputs, yet this is beyond the scope of this work.

\section{Research approach}
We aim at understanding how can better and fairer decisions be taken in refugee status determination.\par

This work focuses on the cognitive process and its consequences on the decision, and does not investigate the institutional procedures which impact the outcome of a decision as well. In the same way that institutions and guidelines provide structures, incentives and frameworks to make accurate judgments, we assume that AI-powered tools could help render noise-free judgments. This hypothesis is supported by the observation that although algorithms do not prevent biases and discrimination, they remove noise. Thereby, we assume that cooperation with machine learning tools and relevant AI functionalities can lead to more justice, in the sense that it can effectively support the decision process of the decision maker. Human-AI cooperation is the proposed way to mitigate risks and errors, combining benefits from AI such as computational power with human abilities such as intuition, background knowledge and context-aware reasoning.\par

We propose a methodology to verify this hypothesis on the case of asylum decisions.

\section{Variability in refugee status adjudications}

\subsection{Legal and explicit criteria in asylum decisions}
Explicit criteria for refugee status determination directly derive from the definition introduced by the international convention and protocol, that is to say the claimant's application has to be justified on the ground of one of the following reasons: race, religion, nationality, membership of a particular social group, political opinion \cite{goodwin-gill_refugee_2021}. While the proof standard tends to be low in refugee status determination, it is a well-informed decision that does not only rely on the analysis of an applicant's specific case, but is also supported by objective information such as country reports. \par

Besides those explicit criteria, the most important evaluation criterion is the credibility assessment for which a ``benefit of the doubt" rule operates as a safeguard \cite{kagan_credibility_nodate}. Credibility is particularly difficult to assess for obvious reasons of trauma, cultural differences, languages, and translations. For all these reasons, asylum decisions are an example of a decision taken under uncertainty with sparse data and high-stakes consequences.\par

\subsection{What is a ``good" decision"?}
For the decision maker, the benefit of the doubt rule in asylum decisions is to balance with the suspicion that people could seek asylum under false pretexts, being a potential threat to a country's safety. The minimal requirements for a decision to be considered as correct and fair to an applicant are:
\begin{enumerate}
    \item Consistency with the local law and administrative guidelines and with the international conventions
    \item Consistency with the evaluation of the situation in the country of the applicant as described in country reports
\end{enumerate}

\subsection{Significance of external features in decision predictions}
Variability in asylum decision has been  previously highlighted by researchers \cite{chen_asylum_2017, dunn_early_2017, schoenholzroulette2007refugee, schoenholzroulette2007refugee, rehaag2012judicial} and in mainstream press \cite{NYTroulette}. \par

Evidence of noise have been found in asylum decisions and experimental results show that prediction can be derived from very few features with satisfactory accuracy. Most features used for prediction are non-substantive, non-legal, i.e. external features. We mainly rely on two studies conducted on asylum decisions taken in the United-States. 
\begin{itemize}
    \item Using a random forest model, \citeauthor{chen_asylum_2017} use 137 features to predict the outcome of asylum decisions, divided in six categories: court information (such as location), judge information, news trend, trend features and weather, achieving an accuracy of 82\% in classifying cases. More importantly, they found that case information only accounts for 20\% of the prediction, the remaining 80\% of the weight being external features.
    \item With the aim of demonstrating the importance of external parameters, \citeauthor{dunn_early_2017} achieve an accuracy of 80\% using three features: judge identity, notice of hearing date, nationality of the applicant. Using the same model with one feature only, judge identity, the classifier still achieves an accuracy of 71\%. 
\end{itemize}

This suggests that asylum decisions are largely influenced by external factors and therefore highly subject to noise. Thus, there is evidence that facts, personal story and provided evidence supporting an applicant's case are not the main factor that determines the outcome of a decision.\par

\subsection{Possible explanations} \label{section 3.4}
There are several hypothesis to explain the apparent randomness of asylum decisions that cannot be reduced to the presence of noise or bias and from which we will derive work hypothesis in section \ref{section 5.3}.\par

\begin{enumerate}
    \item \citeauthor{rehaag2007troubling} studied patterns in asylum decisions in Canada and found similar results as the ones outlined above in decisions rendered in the United-States. He demonstrated that the outcome of a case largely varies depending on the identity of the appointed judge, some granting refugee status in 95.9\% of the cases, some 1.5\% of cases \cite{rehaag2012judicial}. It is important to account for administrative reasons that partially explain this gap. For instance some judges are expert in a geographical area or on a specific type of case, two factors that affect their rates in that cases presented to a judge may consistently fall into a category that does not meet criteria for granting refugee status defined by the Geneva convention. However, it is clear that these explanations do not fully account for the variation in the rate of granting asylum.
    \item Second, it is generally admitted that the cognitive capacities of the human brain are limited and that the treatment of information necessary to effective and efficient decision making is affected by an overload of information \cite{gigerenzer1996reasoning, kahneman2013prospect}. More precisely, the human brain has limited resources in terms of computing capacities, attention, focus and memory. Thereby, we believe that outsourcing some of the workload using AI functionalities should allow for easier information treatment and therefore lead to better decisions.
    \item We make a third hypothesis specific to the case of asylum decisions, that should be verified in future work. We consider that decision makers are likely to take ``short-cuts" in evaluating a case. That is to say that an adjudicator may rely mostly on country reports to decide on granting asylum or not rather than assessing facts and evidence specific to a case. 
\end{enumerate}

\section{Decision process and legal reasoning}

Looking at ``augmenting" human reasoning using data-driven tools would first require to breakdown the human reasoning process into machine-understandable steps. Ideally, this amounts to finding logical steps that can be replicated in the form of inferences and causal links. However, this is not an easy process since empirical evidence shows that decisions are oftentimes incompatible with logic. Nonetheless, we can suggest two paths commonly used in legal decision making.\par

First, legal reasoning can be based on evaluation of alternatives \cite{sep-legal-reas-interpret}, which from a computational point of view would amount to measuring and comparing the risk of not granting asylum to an applicant who fully qualifies against the risk of granting it to someone who does not qualify with regard to the law.

A second approach is to think in terms or past cases and precedent \cite{seplamond-legal-reas-prec, weinreb_legalanalogy_2016} i.e. to trust the experience of the decision maker and its capacity to retrieve and refer to relevant case law.\par 

We assume that legal reasoning in refugee law can be based on analogy. That is to say a new case should be treated based on how similar cases have been treated in the past, which justifies the use of predictive analysis and retrieval of past cases based on similarity, commonly referred to as case-based reasoning.

\section{Potential benefits of data-driven tools}
\subsection{Existing functionalities in legal AI} \label{section 5.1}

Common functionalities in legal AI are:
\begin{itemize}
    \item Legal prediction
    \item Database management and organization
    \item Information retrieval
    \item Similarity analysis and past cases retrieval
    \item Comparison and matching with legal texts and documentation (e.g. legislation, country reports)
    \item Summarization of legislation or cases
\end{itemize}

\subsection{How can noise be measured?}
Measuring noise requires multiple data points, that is to say to think statistically about a set of decisions. 
\begin{enumerate}
    \item interpersonal noise (level and pattern noise) -- by comparing decisions of judges with decisions other judges would have taken on the same case and identifying which features trigger their decision.
    \item intrapersonal noise (occasion noise) -- by asking judges to decide on a case a second time to check if they agree with themselves on a second evaluation, and by comparing a decision with the criteria justifying its outcome.
\end{enumerate}

\subsection{How machine learning functionalities could help render noise-free decisions: proposed experiments} \label{section 5.3}
Building on functionalities listed in section \ref{section 5.1}, we propose to pursue data-driven decision support based on the following links between functionalities and noise reduction in asylum adjudications. Specifically, we aim at proposing solutions to build awareness, detect the magnitude of external features in decision outcomes, and to answer explanations listed in section \ref{section 3.4}.
\begin{itemize}
    \item Evaluate magnitude of external features: (1) Prediction of the outcome to detect features that impact the decision the most and detecting patterns in judges identity, reasons for applications, country of origin. (2) Keywords and argument analysis, which should help understanding the reasons for a decision's outcome and should help in drafting applications.
    
    \item Complement human cognitive capacities: (1) Database organization and management for legislation to reduce the workload and facilitate access to relevant texts. (2) Summarization of the applications (text application and hearing transcription) to detect and extract arguments that are relevant to make a decision (on what justification is the application submitted to the court for instance). (3) Similar past cases retrieval, on criteria such as similar facts and reason for application, country of origin, which should help adjudicators to make fairer decision in the sense that two similar cases should trigger the same decision outcome. 
    
    \item Evaluate the importance of country reports: match cases with country reports in order to check our hypothesis that country of origin is one of the main factors impacting the decisions. Verifying the hypothesis would amount to check the correlation between (1) the country of origin of the applicant and its claimed reason for requesting protection, (2) the description of the situation of this country and community in the country report and (3) the outcome of the case. 

\end{itemize}

\section*{Conclusion}
It is important to state that we do not think that autonomous decisions rendered by AI tools would benefit the decision process in refugee status adjudication. This is why this work aims at proposing tools to support decision making without removing human judgment and expertise. Introducing human-centered computing and the right balance between AI functionalities and human reasoning will help take a step towards cognitive legal computing where machine and human each perform the kind of reasoning they do best. This approach should naturally generate more trust for claimants and legal professionals. 

\bibliography{references}

\begin{thebibliography}{31}
\providecommand{\natexlab}[1]{#1}
\providecommand{\url}[1]{\texttt{#1}}
\expandafter\ifx\csname urlstyle\endcsname\relax
  \providecommand{\doi}[1]{doi: #1}\else
  \providecommand{\doi}{doi: \begingroup \urlstyle{rm}\Url}\fi

\bibitem[UNHCR(1951)]{unhcr_convention_1951}
UNHCR.
\newblock Convention and {Protocol} {Relating} to the {Status} of {Refugees},
  1951.
\newblock URL
  \url{https://www.unhcr.org/protection/basic/3b66c2aa10/convention-protocol-relating-status-refugees.html}.

\bibitem[UN(1967)]{un_protocol_1967}
UN.
\newblock Protocol relating to the {Status} of {Refugees}, 1967.
\newblock URL
  \url{https://www.ohchr.org/EN/ProfessionalInterest/Pages/ProtocolStatusOfRefugees.aspx}.

\bibitem[Allen(1956)]{allenlayman_symbolic_1956}
Layman~E. Allen.
\newblock Symbolic logic: {A} razor-edged tool for drafting and interpreting
  legal documents.
\newblock \emph{Yale LJ}, 66:\penalty0 833, 1956.
\newblock Publisher: HeinOnline.

\bibitem[Buchanan and Headrick(1970)]{buchananheadrick_speculation_1970}
Bruce~G. Buchanan and Thomas~E. Headrick.
\newblock Some speculation about artificial intelligence and legal reasoning.
\newblock \emph{Stan. L. Rev.}, 23:\penalty0 40, 1970.
\newblock Publisher: HeinOnline.

\bibitem[Popple(1996)]{popple_pragmatic_1996}
James Popple.
\newblock \emph{A pragmatic legal expert system}.
\newblock Dartmouth, Aldershot ; Brookfield, USA, 1996.
\newblock ISBN 978-1-85521-739-3.

\bibitem[Molnar and Gill(2018)]{molnar_botsgate_2018}
Petra Molnar and Lex Gill.
\newblock Bots at the {Gate}: {A} {Human} {Rights} {Analysis} of {Automated}
  {Decision}-{Making} in {Canada}’s {Immigration} and {Refugee} {System}.
\newblock Technical report, University of Toronto, September 2018.

\bibitem[{European Commission. Directorate General for Migration and Home
  Affairs.} and
  {Deloitte.}(2020)]{european_commission_directorate_general_for_migration_and_home_affairs_opportunities_2020}
{European Commission. Directorate General for Migration and Home Affairs.} and
  {Deloitte.}
\newblock \emph{Opportunities and challenges for the use of artificial
  intelligence in border control, migration and security. {Volume} 1, {Main}
  report.}
\newblock Publications Office, LU, 2020.
\newblock URL \url{https://data.europa.eu/doi/10.2837/923610}.

\bibitem[Service(2020)]{service_unhcrmediumgiving_2020}
UNHCR~Innovation Service.
\newblock Giving {Legal} {Teams} {Better} {Tools} to {Represent} {Asylum}
  {Seekers}, June 2020.
\newblock URL
  \url{https://medium.com/unhcr-innovation-service/giving-legal-teams-better-tools-to-represent-asylum-seekers-df7802e815df}.

\bibitem[Cameron et~al.(2021)Cameron, Goldfarb, and
  Morris]{cameron_artificial_2021}
Hilary~Evans Cameron, Avi Goldfarb, and Leah Morris.
\newblock Artificial intelligence for a reduction of false denials in refugee
  claims.
\newblock \emph{Journal of Refugee Studies}, page feab054, May 2021.
\newblock ISSN 0951-6328, 1471-6925.
\newblock \doi{10.1093/jrs/feab054}.
\newblock URL
  \url{https://academic.oup.com/jrs/advance-article/doi/10.1093/jrs/feab054/6271402}.

\bibitem[Dunn et~al.(2017)Dunn, Sagun, Şirin, and Chen]{dunn_early_2017}
Matt Dunn, Levent Sagun, Hale Şirin, and Daniel Chen.
\newblock Early predictability of asylum court decisions.
\newblock In \emph{Proceedings of the 16th edition of the {International}
  {Conference} on {Articial} {Intelligence} and {Law}}, pages 233--236, London
  United Kingdom, June 2017. ACM.
\newblock ISBN 978-1-4503-4891-1.
\newblock \doi{10.1145/3086512.3086537}.
\newblock URL \url{https://dl.acm.org/doi/10.1145/3086512.3086537}.

\bibitem[Chen and Eagel(2017)]{chen_asylum_2017}
Daniel~L. Chen and Jess Eagel.
\newblock Can machine learning help predict the outcome of asylum
  adjudications?
\newblock In \emph{Proceedings of the 16th edition of the {International}
  {Conference} on {Articial} {Intelligence} and {Law}}, pages 237--240, London
  United Kingdom, June 2017. ACM.
\newblock ISBN 978-1-4503-4891-1.
\newblock \doi{10.1145/3086512.3086538}.
\newblock URL \url{https://dl.acm.org/doi/10.1145/3086512.3086538}.

\bibitem[Katz et~al.(2017)Katz, Bommarito, and Blackman]{katz_general_2017}
Daniel~Martin Katz, Michael~J. Bommarito, and Josh Blackman.
\newblock A general approach for predicting the behavior of the {Supreme}
  {Court} of the {United} {States}.
\newblock \emph{Plos one}, 12\penalty0 (4):\penalty0 e0174698, April 2017.
\newblock ISSN 1932-6203.
\newblock \doi{10.1371/journal.pone.0174698}.
\newblock URL \url{https://dx.plos.org/10.1371/journal.pone.0174698}.

\bibitem[Martin et~al.(2004)Martin, Quinn, Ruger, and
  Kim]{martin_competingsupreme_2004}
Andrew~D. Martin, Kevin~M. Quinn, Theodore~W. Ruger, and Pauline~T. Kim.
\newblock Competing {Approaches} to {Predicting} {Supreme} {Court} {Decision}
  {Making}.
\newblock \emph{Perspectives on Politics}, 2\penalty0 (04):\penalty0 761--767,
  December 2004.
\newblock ISSN 1537-5927, 1541-0986.
\newblock \doi{10.1017/S1537592704040502}.
\newblock URL
  \url{http://www.journals.cambridge.org/abstract_S1537592704040502}.

\bibitem[Ruger et~al.(2004)Ruger, Kim, Martin, and
  Quinn]{ruger_supremeforecast_2004}
Theodore~W. Ruger, Pauline~T. Kim, Andrew~D. Martin, and Kevin~M. Quinn.
\newblock The {Supreme} {Court} {Forecasting} {Project}: {Legal} and
  {Political} {Science} {Approaches} to {Predicting} {Supreme} {Court}
  {Decisionmaking}.
\newblock \emph{Columbia Law Review}, 104\penalty0 (4):\penalty0 1150, May
  2004.
\newblock ISSN 00101958.
\newblock \doi{10.2307/4099370}.
\newblock URL \url{https://www.jstor.org/stable/4099370?origin=crossref}.

\bibitem[Undavia et~al.(2018)Undavia, Meyers, and
  Ortega]{undavia_comparative_2018}
Samir Undavia, Adam Meyers, and John Ortega.
\newblock A {Comparative} {Study} of {Classifying} {Legal} {Documents} with
  {Neural} {Networks}.
\newblock pages 515--522, September 2018.
\newblock \doi{10.15439/2018F227}.
\newblock URL \url{https://fedcsis.org/proceedings/2018/drp/227.html}.

\bibitem[Aletras et~al.(2016)Aletras, Tsarapatsanis, Preoţiuc-Pietro, and
  Lampos]{aletras_predicting_2016}
Nikolaos Aletras, Dimitrios Tsarapatsanis, Daniel Preoţiuc-Pietro, and
  Vasileios Lampos.
\newblock Predicting judicial decisions of the {European} {Court} of {Human}
  {Rights}: a {Natural} {Language} {Processing} perspective.
\newblock \emph{PeerJ Computer Science}, 2:\penalty0 e93, October 2016.
\newblock ISSN 2376-5992.
\newblock \doi{10.7717/peerj-cs.93}.
\newblock URL \url{https://peerj.com/articles/cs-93}.

\bibitem[Medvedeva et~al.(2020)Medvedeva, Vols, and
  Wieling]{medvedeva2020using}
Masha Medvedeva, Michel Vols, and Martijn Wieling.
\newblock Using machine learning to predict decisions of the european court of
  human rights.
\newblock \emph{Artificial Intelligence and Law}, 28\penalty0 (2):\penalty0
  237--266, 2020.

\bibitem[Kaur and Bozic(2019)]{kaur2019convolutional}
Arshdeep Kaur and Bojan Bozic.
\newblock Convolutional neural network-based automatic prediction of judgments
  of the european court of human rights.
\newblock In \emph{AICS}, pages 458--469, 2019.

\bibitem[Rehaag(2012)]{rehaag2012judicial}
Sean Rehaag.
\newblock Judicial review of refugee determinations: The luck of the draw.
\newblock \emph{Queen's LJ}, 38:\penalty0 1, 2012.

\bibitem[Daniel~Kahneman(2021)]{noisekahneman}
Cass R.~Sunstein Daniel~Kahneman, Olivier~Sibony.
\newblock \emph{Noise : a flaw in human judgement}.
\newblock William Collins, London, 2021.
\newblock ISBN 9780008308995.

\bibitem[Sunstein(2021)]{sunstein2021governing}
Cass~R Sunstein.
\newblock Governing by algorithm? no noise and (potentially) less bias.
\newblock \emph{No Noise and (Potentially) Less Bias (September 15, 2021)},
  2021.

\bibitem[Goodwin-Gill and McAdam(2021)]{goodwin-gill_refugee_2021}
Guy~S. Goodwin-Gill and Jane McAdam.
\newblock \emph{The refugee in international law}.
\newblock Oxford University Press, New York, fourth edition edition, 2021.
\newblock ISBN 978-0-19-880857-2 978-0-19-880856-5.

\bibitem[Kagan()]{kagan_credibility_nodate}
Michael Kagan.
\newblock Is truth in the eye of the beholder? {Objective} credibility
  assessment in refugee status determination.
\newblock \emph{Georgetown immigration law journal}, 17:\penalty0 51.

\bibitem[Ramji-Nogales et~al.(2007)Ramji-Nogales, Schoenholtz, and
  Schrag]{schoenholzroulette2007refugee}
Jaya Ramji-Nogales, Andrew~I Schoenholtz, and Philip~G Schrag.
\newblock Refugee roulette: Disparities in asylum adjudication.
\newblock \emph{Stan. L. Rev.}, 60:\penalty0 295, 2007.

\bibitem[Bernstein(2006)]{NYTroulette}
Nina Bernstein.
\newblock In new york immigration court, asylum roulette.
\newblock \emph{The New York Times}, 2006.

\bibitem[Rehaag(2007)]{rehaag2007troubling}
Sean Rehaag.
\newblock Troubling patterns in canadian refugee adjudication.
\newblock \emph{Ottawa L. Rev.}, 39:\penalty0 335, 2007.

\bibitem[Gigerenzer and Goldstein(1996)]{gigerenzer1996reasoning}
Gerd Gigerenzer and Daniel~G Goldstein.
\newblock Reasoning the fast and frugal way: models of bounded rationality.
\newblock \emph{Psychological review}, 103\penalty0 (4):\penalty0 650, 1996.

\bibitem[Kahneman and Tversky(2013)]{kahneman2013prospect}
Daniel Kahneman and Amos Tversky.
\newblock Prospect theory: An analysis of decision under risk.
\newblock In \emph{Handbook of the fundamentals of financial decision making:
  Part I}, pages 99--127. World Scientific, 2013.

\bibitem[Dickson(2016)]{sep-legal-reas-interpret}
Julie Dickson.
\newblock {Interpretation and Coherence in Legal Reasoning}.
\newblock In Edward~N. Zalta, editor, \emph{The {Stanford} Encyclopedia of
  Philosophy}. Metaphysics Research Lab, Stanford University, {W}inter 2016
  edition, 2016.

\bibitem[Lamond(2016)]{seplamond-legal-reas-prec}
Grant Lamond.
\newblock {Precedent and Analogy in Legal Reasoning}.
\newblock In Edward~N. Zalta, editor, \emph{The {Stanford} Encyclopedia of
  Philosophy}. Metaphysics Research Lab, Stanford University, {S}pring 2016
  edition, 2016.

\bibitem[Weinreb(2016)]{weinreb_legalanalogy_2016}
Lloyd~L. Weinreb.
\newblock \emph{Legal reason : the use of analogy in legal argument / {Lloyd}
  {L}. {Weinreb}.}
\newblock Cambridge University Press, New York, NY, USA, second edition.
  edition, 2016.
\newblock ISBN 978-1-316-60732-9.
\newblock Publication Title: Legal reason : the use of analogy in legal
  argument.

\end{thebibliography}
\bibliographystyle{unsrtnat}
\setcitestyle{authoryear,open={((},close={))}}

\end{multicols}

\end{document}